\documentclass[12pt,aps,prb,preprint]{revtex4}
\usepackage{amsmath} 
\begin{document}

\title{What is the sign of $\hbar$ ?}
\author{Massimo Testa}
 \affiliation{Dipartimento di Fisica, Universit\`a di Roma ``La
Sapienza",  Sezione INFN di Roma\\ P.le A. Moro 2, 00185 Roma, Italy}

\begin{abstract}
We present an elementary argument showing that the sign of $\hbar$ in the basic formulation of Quantum Mechanics can be changed without incurring in any physical consequences.
\end{abstract}
\maketitle
\section{Canonical Quantization}

Poisson brackets
\begin{equation}
\{f(q,p),g(q,p)\} \equiv \sum_{i=1}^d \left ( {{\partial f} \over {\partial q_i}} { {\partial g} \over {\partial p_i}} -  {{\partial g} \over {\partial q_i}} { {\partial f} \over {\partial p_i}}   \right ) \label{poisson}
\end{equation}
are introduced in Classical Mechanics as a sophisticated instrument to deal with canonical transformations\cite{gold}. In Quantum Mechanics, on the contrary, they are the basic objects in the Dirac's\cite{dirac} quantization method, which is based on some common properties between operator commutators, $\left [f(q,p),g(q,p)\right ]$, and Poisson brackets, eq.(\ref{poisson}):
\begin{itemize}
\item antisymmetry 
\begin{eqnarray}
\{f(q,p),g(q,p) \} = -\{g(q,p),f(q,p) \} \nonumber \\ 
\left [ f(q,p),g(q,p) \right ] = -\left [g(q,p),f(q,p) \right] \nonumber
\end{eqnarray}
\item linearity
\begin{eqnarray}
\{c_1 f_1(q,p)+c_2 f_2(q,p),g(q,p) \} = c_1 \{ f_1(q,p),g(q,p) \} + c_2 \{ f_2(q,p),g(q,p) \} \nonumber \\ 
\left [c_1 f_1(q,p)+c_2 f_2(q,p),g(q,p) \right ] = c_1 \left [f_1(q,p),g(q,p)\right ] + c_2 \left [ f_2(q,p),g(q,p) \right ] \nonumber
\end{eqnarray}
\item Jacobi identity
\begin{eqnarray}
\{f,\{g,h\} \} + \{h,\{f,g\} \} + \{g,\{h,f\} \} =0 \nonumber \\
\left [f,\left [g,h \right ] \right ] + \left [h,\left [f,g\right ] \right ] + \left [g,\left [h,f\right ]\right ] =0 \, . \nonumber
\end{eqnarray}
\end{itemize}
On the basis of these analogies Dirac concludes that commutators and Poisson brackets must be proportional
\begin{equation}
\left [ f(q,p),g(q,p) \right ] = i \hbar \{f(q,p),g(q,p)\} \, . \label{quant}
\end{equation}
The factor of $i$ in eq.(\ref{quant}) is due to the anti-hermiticity of the commutator of hermitian operators and $\hbar$ is, according to Dirac, a parameter which must be fixed by experiment. $\hbar$ has the dimensions of an action in virtue of the definition of Poisson bracket and its value is experimentally found to be
\begin{equation}
\hbar= 1.054 \cdot10^{-34} \text{Joule} \cdot \text{sec} \, . \label{hbar}
\end{equation}
$\hbar$ in eq.(\ref{quant}) has been tacitly {\it assumed} to be positive, so that the question naturally arises if any experiment can fix its sign. In other words, if we continue to denote by $\hbar$ the value given in eq.(\ref{hbar}), the question is if the correct quantization condition is eq.(\ref{quant}) or rather
\begin{equation}
\left [ f(q,p),g(q,p) \right ] = - i \hbar \{f(q,p),g(q,p)\} \, . \label{quant1}
\end{equation}
This question is legitimate because, starting from eq.(\ref{quant}), when applied to position and momentum operators\cite{1dim},
\begin{equation}
\left [q,p \right ] = i \hbar \label{cancomm}
\end{equation}
we get the famous identification of the momentum operator
\begin{equation}
p = - i \hbar {{d }\over {d x}} \, , \label{mom}
\end{equation}
while, starting from eq.(\ref{quant1}), the opposite sign would be obtained in eq.(\ref{mom}). Would this contrast some empirical data?

It is well known\cite{dirac} that predictions of Quantum Mechanics are not changed if we apply an arbitrary unitary transformation $U$ to states and operators, because in this way all matrix elements remain invariant
\begin{equation}
\langle \phi | O | \psi \rangle = \langle \phi | U^\dagger (U O U^\dagger) U | \psi \rangle \, .
\end{equation}
It is also clear that, since unitary transformations do not alter the structure of commutation relations, the theory with a negative $\hbar$ cannot be unitarily equivalent to the original one.

It is the purpose of this paper to show that the change in sign of $\hbar$ is equivalent to an anti-unitary transformation which, as we shall argue, leaves physics unchanged. 

\section{Antilinear operators}

In this section we will discuss some concepts that, although well known, are not usually discussed in introductory courses in Quantum Mechanics. The simplest way to discuss antilinear operators is to give up the Dirac formalism and work directly with wave functions, $\psi (x)$.
An operator ${\cal A}$ is called antilinear if
\begin{equation}
{\cal A} (c_1 \psi_1+ c_2 \psi_2 ) =  c^*_1 {\cal A} \psi_1+ c^*_2 {\cal A}  \psi_2 \, .
\end{equation}
The vector space of the $\psi$'s is  endowed with the scalar product
\begin{equation}
(\phi,\psi) \equiv \int_{-\infty}^\infty \phi^*(x) \psi(x) \, dx
\end{equation}
which is linear in $\psi$ and antilinear in $\phi$.
In this formalism the hermitian conjugate of any operator $O$ is defined through
\begin{equation}
(\phi,O \psi) = (O^\dagger \phi,\psi) \, . \label{herm}
\end{equation}
$O^\dagger$ is a linear operator because only in this case both sides of eq.(\ref{herm}) are linear in $\psi$ and antilinear in $\phi$.
In the case of antilinear operators the definition of ${\cal A}^\dagger$ has to be modified as
\begin{equation}
(\phi,{\cal A}  \psi) = ({\cal A} ^\dagger \phi,\psi)^* \, , \label{antiherm}
\end{equation}
because in this way ${\cal A} ^\dagger$ is antilinear and both sides of eq.(\ref{antiherm}) are antilinear in $\psi$ and $\phi$.

\section{Changing the sign of $\hbar$}

We now proceed to show that changing the sign of $\hbar$ is equivalent to perform on states and observables the simplest antiunitary transformation ${\cal K}$, defined as\cite{wigner}
\begin{equation}
{\cal K} \psi(x) \equiv  \psi_{\cal K} (x) = \psi^*(x) \, . \label{defk}
\end{equation}
Let us check that ${\cal K}$ is anti-unitary
\begin{equation}
({\cal K} \phi , {\cal K} \psi)=  \int_{-\infty}^\infty \phi (x) \psi^* (x) \, dx =(\psi ,\phi) = ({\cal K}^\dagger  {\cal K} \phi , \psi)^*= (\psi, {\cal K}^\dagger  {\cal K} \phi) \, ,
\end{equation}
from which we conclude
\begin{equation}
{\cal K}^\dagger  {\cal K} = I \, .
\end{equation}
Eq.(\ref{defk}) also implies
\begin{eqnarray}
{\cal K} x \psi (x) = x \psi^* (x) = {\cal K} x {\cal K}^\dagger  {\cal K} \psi (x) = {\cal K} x {\cal K}^\dagger \psi^* (x) 
\end{eqnarray}
so that
\begin{equation}
{\cal K} x {\cal K}^\dagger \equiv x_{\cal K}= x \label{kpos}
\end{equation}
and
\begin{eqnarray}
{\cal K} p \psi (x) = {\cal K} (-i \hbar) {{d \psi (x)} \over {d x}}= i \hbar {{d \psi^* (x)} \over {d x}}
= {\cal K} p {\cal K}^\dagger  {\cal K} \psi (x) = {\cal K} p {\cal K}^\dagger \psi^* (x) 
\end{eqnarray}
so that
\begin{equation}
{\cal K} p {\cal K}^\dagger \equiv p_{\cal K} = i \hbar {{d} \over {d x}} = -p \, .\label{kmom}
\end{equation}
Eqs.(\ref{kpos},\ref{kmom}) are consistent with the ${\cal K}$ transform of eq.(\ref{cancomm})
\begin{equation}
 {\cal K} \left [q,p \right ]  {\cal K}^\dagger=  \left [q_{\cal K} ,p_{\cal K} \right ] = {\cal K} i \hbar {\cal K}^\dagger = - i \hbar
  \label{cancomm1}
\end{equation}

Eqs.(\ref{kpos},\ref{kmom},\ref{cancomm1}) imply that the change of the sign of $\hbar$ is equivalent to transform the whole theory under ${\cal K}$. It must be remarked that the Schr{\oe}dinger equation
\begin{equation}
i \hbar { {\partial \psi} \over{\partial t}} =H \psi
\end{equation}
is modified by the ${\cal K}$ transformation as
\begin{equation}
- i \hbar { {\partial \psi_{\cal K}} \over{\partial t}} =H_{\cal K} \psi_{\cal K} \, .
\end{equation}
In particular, for a particle moving in an external electromagnetic field described by a vector potential ${\bf A}({\bf x}, t)$ and a scalar potential $\Phi({\bf x}, t)$, we have
\begin{equation}
H=  {{1} \over {2m}} ( i \hbar \boldsymbol{\nabla} + {{e} \over {c}} {\bf A}({\bf x}, t) )^2 + e \Phi ({\bf x}, t)
\end{equation}
and
\begin{equation}
H_{\cal K} =  {{1} \over {2m}} (- i \hbar \boldsymbol {\nabla} + {{e} \over {c}} {\bf A}({\bf x}, t) )^2 + e \Phi ({\bf x}, t) \, .
\end{equation}
The changes induced by the substitution of $\hbar$ into $- \hbar$ look non trivial; it is however easy to see how the matrix elements of any observable (hermitian) operator $O$ ($= O^\dagger$) behave under this transformation. We have
\begin{eqnarray}
(\phi, O \psi) = ({\cal K} \phi, {\cal K} O \psi)^* = ({\cal K} \phi, {\cal K} O {\cal K}^\dagger {\cal K}\psi)^* =
(\phi_{\cal K}, O_{\cal K} \psi_{\cal K})^* \, . \label{matrix}
\end{eqnarray}
Eq.(\ref{matrix}) shows that inverting the sign of $\hbar$ in eq.(\ref{quant}) amounts to change the matrix elements of all observables into their complex conjugates. This implies that quantum averages (real diagonal matrix elements) are invariant, while off-diagonal matrix elements are not. However only absolute values of off-diagonal matrix elements are physically measurable and therefore we can safely conclude that altering the sign of $\hbar$ does not lead to any observable effect.

In the spinless case $\psi_{\cal K}$ describes the state with time reversed properties\cite{wigner} with respect to $\psi$. In fact eqs.(\ref{kpos}) and (\ref{kmom}) give
\begin{eqnarray}
(\psi_{\cal K} , x \, \psi_{\cal K})= (\psi , x \, \psi) \nonumber \\
(\psi_{\cal K} , p \, \psi_{\cal K})= - (\psi , p \, \psi) \nonumber \, , 
\end{eqnarray}
showing that $\psi_{\cal K}$ is a state with unchanged average position and reversed average momentum, i.e. it describes the state time-reversed with respect to $\psi$. I want to stress that, in spite of this,
time reversal invariance is not required in the proof of the invariance of physical results under the substitution $\hbar\rightarrow - \hbar$: in fact in eq.(\ref{matrix}) {\it both} states and operators are transformed under $\cal K$. The ${\cal K}$ transformation used in this way only reflects a redundancy of the formalism which does not change the values of observables. Moreover time reversal invariance would impose the additional restriction on dynamics $H_{\cal K} = H$, which was never required.

The transformation ${\cal K}$ is just the simplest antiunitary transformation: it is easy to go through the arguments presented in the previous sections and realise that {\it any} antiunitary operator would have the same effect of changing the sign of $\hbar$ in the basic algebra, eq.(\ref{quant}), while leaving physics untouched. It is however true that the time reversal operator performs this job in a more aesthetical way. Let me exemplify this fact through the case of a spin-1/2 particle. The spin degrees of freedom, ${\bf s}$, are described through the Pauli matrices $\boldsymbol {\sigma}$ as
\begin{equation}
{\bf s} = {{\hbar}\over {2}} \boldsymbol {\sigma}  \label{spinop}
\end{equation}
and the time reversal operator, $\Theta$, in this case is\cite{wigner}
\begin{equation}
\Theta = \sigma _y \, {\cal K} \, .
\end{equation}
While ${\cal K}$ and $\Theta$ act on orbital degrees of freedom in the same way, they transform ${\bf s}$ quite differently. Under ${\cal K}$ we have
\begin{eqnarray}
s_{{\cal K} x} \equiv {\cal K} s_x {\cal K} ^\dagger = s_x \nonumber \\
s_{{\cal K} y} \equiv {\cal K} s_y {\cal K} ^\dagger = - s_y \label{spintransf} \\
s_{{\cal K} z} \equiv {\cal K} s_z {\cal K} ^\dagger = s_z \nonumber \, ,
\end{eqnarray}
while, under $\Theta$,
\begin{equation}
{\bf s}_{\Theta} \equiv \Theta {\bf s} \Theta ^\dagger = - {\bf s} \label{spintime} \, .
\end{equation}
Both eqs.(\ref{spintransf}) and (\ref{spintime}) provide sets of spin operators obeying commutation rules with $\hbar \rightarrow - \hbar$
\begin{eqnarray*}
[ s_{{\cal K} x}, s_{{\cal K} y}] = -i \hbar s_{{\cal K} z} \\
\end{eqnarray*}
\begin{equation*}
[ s_{\Theta x}, s_{\Theta y}] = -i \hbar s_{\Theta z}
\end{equation*}
for which the argument given in eq.(\ref{matrix}) is valid. However only the $\Theta$ transformation corresponds to change $\hbar \rightarrow - \hbar$ also inside the operator expression, eq.(\ref{spinop}): the choice between ${\cal K}$ and $\Theta$ is only a matter of taste. Similar arguments are also valid in quantum field 
theory.


\begin{thebibliography}{99}
\bibitem{1dim} We consider the one dimensional case in order to simplify the notations.
\end{thebibliography}
\end{document}